# In situ Measurement of Airborne Particle Concentration in a Real Dental Office: Implications for Disease Transmission

Maryam Ravazi[1], Zahid Butt[2], Mark H.E. Lin[3], Helen Chen[2], Zhongchao Tan[1*]

[1] Department of Mechanical and Mechatronics Engineering, University of Waterloo, Ontario, Canada

[2] School of Public Health and Health Systems, University of Waterloo

[3] The Institute for Dental Excellence Inc., 88 Finch Ave, East North York, Ontario, Canada

[*] tanz@uwaterloo.ca, Phone: 001 226-929-8021

**Abstract**

Recent guidelines by WHO recommend delaying non-essential oral health care amid COVID-19 pandemic and call for research on aerosol generated during dental procedures. Thus, this study aims to assess the mechanisms of dental aerosol dispersion in dental offices and to provide recommendations based on a quantitative study to minimize infection transmission in dental offices. The spread and removal of aerosol particles generated from dental procedures in a dental office are measured near the source and at the corner of the office. We studied the effects of air purification (on/off), door condition (open/close), and particle sizes on the temporal concentration distribution of particles. The results show that in the worst-scenario scenario it takes 95 min for 0.5 $\mu$m particles to settle, and that it takes a shorter time for the larger particles. The indoor air purifier tested expedited the removal time at least 6.3 times faster than the scenario air purifier off. Airborne particles may be transported from the source to the rest of the room, even when the particle concentrations in the generation zone return to the background level. These results are expected to be valuable to related policy making and technology development for infection disease control in dental offices and similar built environments.

**Highlights:**

- Quantification of the time it takes for dental aerosol concentration to return to the background level.

- Identification of the best practices for a dental office in COVID-19 pandemic.

- Dependence of the temporal distribution of aerosols generated during dental procedures on air purification and ventilation.

## 1. Introduction

In August 2020, the world health organization (WHO) guidelines recommend delaying routine non-essential oral health care amid COVID-19 pandemic and call for more research on indoor aerosol generated by dental procedures. The reason is that dental professionals, staff, and patients in dental offices are exposed to aerosol droplets, particles, and pathogenic microorganisms in the saliva and blood of the infected patients. The infectious microorganisms transmitted from saliva and nasopharyngeal secretions include pneumonic plague, Legionella pneumophila, tuberculosis, influenza viruses, herpes viruses, SARS virus (a form of coronavirus), pathogenic streptococci and staphylococci, HIV, and hepatitis viruses [1, 2]. Recently, COVID-19 joins this list because studies show that dentists are at high risk of exposure to this virus, even more than nurses [3]. These infectious diseases, particularly COVID-19, could be transmitted from pre-symptomatic or asymptomatic patients in the recovery phase [4].

Infectious microorganisms spread in dental offices via various routes [5]. These routes include direct contact with body fluid of infected person, contact with surfaces and instruments that are touched, contact with the exhaled air by the infected person, and infection transmission through aerosols generated during the dental procedures [2, 5-8]. The most considerable one is associated with aerosol smaller than 5 $\mu$m in diameter, recognized by the WHO in healthcare settings [9]. Splatters are another potential source of infection. Splatters are a mixture of air, water, and solid substances [1]. As the water evaporates, the smaller splatters linger longer in the air. In addition, exposure to non-biological aerosol particles in the dental offices and laboratories adversely affects human health [2, 10]. Many researchers have reported similar detrimental effects of the dental aerosol [11-14]. There is a direct correlation between the respiratory system infections of dental personnel and the concentration of generated aerosols due to the dental procedures. Particles deposit in the alveolar region of the respiratory system can further enter the bloodstream, causing lung cancer, pulmonary and cardiovascular diseases, heart diseases, asthma, increased mortality, *etc.* [15, 16].

According to the American Dental Association (ADA), the Centers for Disease Control and Prevention (CDC), and the Occupational Safety and Health Administration (OSHA), all the contaminated aerosol and splatters should be eliminated as much as possible from the air in the dental offices and related laboratories [18, 19]. It is necessary not only for the protection of people in the dental offices but also for the prevention and control of disease transmission. Aerosol particles with a diameter of 50 $\mu$m remain suspended in the air for up to 30 minutes after their formation [20], while smaller particles may remain airborne much longer.

Latest research about COVID-19 suggests the potential for airborne transmission of SARS-CoV-2 (the coronavirus that causes COVID-19) through aerosols [8]. Therefore, strict and effective infection control protocols are highly required to fight COVID-19 in dentaries [5] as well as other indoor spaces. General preventive measures and dental practice recommendations have been developed during the COVID-19 pandemic [21, 22]. The Ontario Dental Association guidelines, for example, require three hours between two patients during the COVID-19 pandemic [23]. These guidelines pose a great challenge to the dental business because of reduced or no patient visits. Systematic research is urgently needed for the development of alternative approaches for the decision-makers.

Protection methods are constantly emphasized in guidelines. Multiple approaches may help reduce the transmission of infectious diseases. The use of personal protection such as facemasks, gloves, and goggles are recommended to reduce the exposure of dental staff to aerosol; however, facemasks are not 100% effective [24, 25]. Measurements show a very high concentration of particles $9.7 \times 10^5$ #/$cm^3$ even behind surgical masks [25]. Rubber dams and suction tubes can protect patients, while their uses are limitted to certain dental operations [2, 25]. High-efficiency particulate air (HEPA) filter, ultraviolet (UV) light, and chambers in ventilation system are other protective methods that are effective after threats have become airborne and spread to the room. Extraoral high volume evacuators (EHVE) can also be used to remove the aerosol particles near to the area of particle generation [1]; however, its performance depends on the volumetric rate of evacuation and particle generation rate. In addition, using extra devices around the dental unit causes a restricted environment and inconvenience to the dentists. Recent COVID-19 outbreak has resulted in increased use of portable air purifiers in dental offices, despite the scarcity of published research on their performances in dental offices [26, 27]. Further research on the protective effectiveness of air purifiers in dental clinics was recommended [28]. The portable air purifiers can be located at the corners in the dental offices, and they cause much less inconvenience during dental operations than extra-oral high evacuators do. In addition, these portable air purifiers do not require modification to existing ventilation systems.

Despite earlier research on number concentrations for micron [29, 30] and nano-size particles [31-33] related to the dental processes, to the best of our knowledge, no research has been done on the dispersion or transport of airborne particles lingering in different parts of the office. The nature of the extensive surface area in dental offices may enhance the losses of particles onto various surfaces. Furthermore, research on the effects of air purifiers is needed to develop guidelines and protocols to reduce waiting time between patients and ensure the safe operation of dental offices.

The objective of this study is to understand the spatial and temporal concentration-distribution of airborne particles generated from dental procedures in dental offices. The remainder of this paper is presented as follows. Section 2 presents the experimental design of concentration measurements in the dental office. Section 3.1 reports the number concentration distribution of particle under the effects of operating conditions during the generation; Section 3.2, the spatial and temporal change of particle concentrations distribution under the effects of operating conditions at the generation zone; Section 3.3, at the corner of the office. The results reveal the effective removal mechanisms that depend on particle size. Finally, Section 5 summarizes the entire work. Results in this paper are deemed valuable to the best practices for particle removal from dental offices.

## 2. Materials and Methods

### 2.1. Measurement site and instruments

The concentrations of micron and submicron particles were measured on May 15, 2020 in a dental operation room on the second floor of the dental clinic in Toronto, Ontario, Canada. Figure 1 shows the schematic of the operatory and layout of the instruments. This typical dental operatory room is 3 m wide, 3 m long, and 4 m high; it has one central dental unit. The mechanical ventilation system was turned off and the window was closed throughout the test. The temperature and relative humidity of the room air were 13.4$^0$C and 88%, respectively.

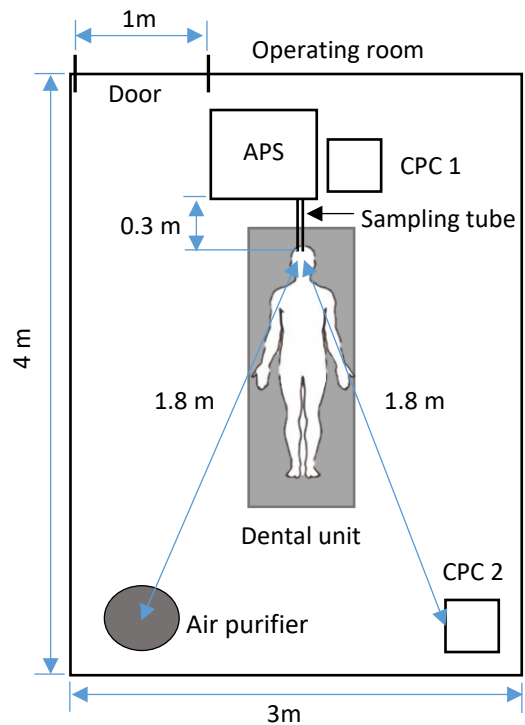

Figure 1. Schematic of the experimental setup

The number concentrations of particles were measured using an aerodynamic particle sizer spectrometer (APS, TSI 3321) and two optical particle counters (OPC, Handheld 3016, Lighthouse Worldwide Solutions Inc.) The APS took data every 5 min with 5 scans; each scan lasted 20 s; it can detect the particles in the range of 0.5-20 µm in diameter and those smaller than 0.5 µm. The APS was located on the left-hand side of the doctor, to prevent any inconvenience for the doctor during dental operations. A stainless-steel sampling tube, which is 1/4-inch of inner diameter and 0.3 m long, was connected to the inlet of the APS for sampling air 10 cm away from the operation area (*i.e*., the patient's mouth). Both OPCs were running continuously. One OPC was located beside the APS, and another OPC was 1.8 m away from the source. Both OPCs report particles with diameters of 0.3, 0.5, 1, 2, 5, 10 µm. The first OPC is calibrated against the APS.

## 2.3. Study design of dental operation on pig jaw

Before the operation, the room was unoccupied for 15 hours before the background concentrations were measured at the source without air purification. As seen in Figure 2, all particles in the background air were less than 10 #/cm$^3$ and those larger than 1 $\mu$m in diameter were less than 1 #/cm$^3$.

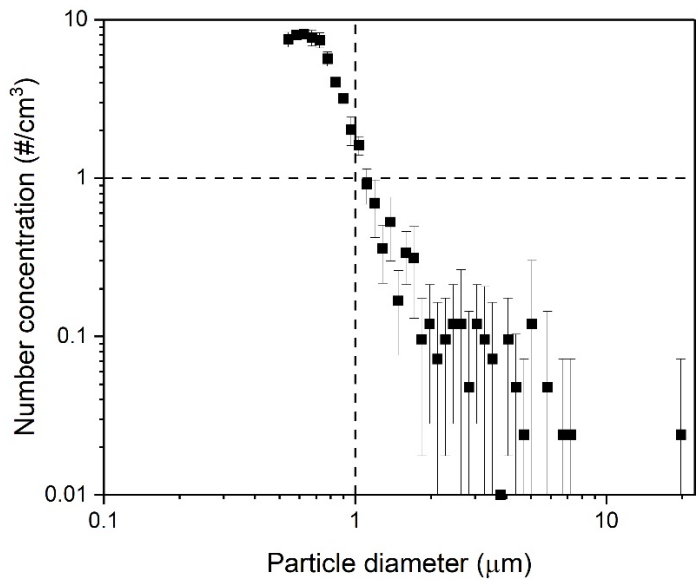

Figure 2. Background number concentration for 0.5- 20 µm aerosols

Airborne particles were generated over 5 min of continuous drilling operation (high-speed handpiece) using a pig jaw. Pig teeth are commonly used for dental studies because of similarities between the structure of human and pig enamel and dentin [34, 35]. The particle number concentrations were measured during 5 min of continuous dental operation and afterward until the number concentrations reached the background. Then we measured the airborne particle concentrations under six scenarios. Table 1 shows the conditions of these scenarios. All particles were generated by drilling the pig jaw with a high-speed handpiece. Other factors considered include the door condition (open and close) and air purifier (on and off, airflow rate, starting time). The air purifier (surgically clean air, model: JADE, SCA5000C) was 1.8 m away from the generation zone.

Table 1. Test scenarios and conditions

| Scenario No. | Dental operation duration | Door (Open/Close) | Air purifier | | |
|---|---|---|---|---|---|
| | | | On/Off | Fan speed | Air cleaning starting time |
| 1 | 5 min | Open | Off | - | - |
| 2 | 5 min | Open | On | High (312 CFM) | At the beginning of the operation |
| 3 | 5 min | Close | Off | - | - |
| 4 | 5 min | Close | On | Low (153 CFM) | After 5 min of operation |
| 5 | 5 min | Close | On | High (312 CFM) | After 5 min of operation |
| 6 | 5 min | Close | On | High (312 CFM) | At the beginning of the operation |

## 3. Results and discussion

### 3.1 Particle generation during operation for five minutes

Figure 3 shows the incremental concentrations, which are defined as the differences between real concentrations and the background, during the five min of continuous dental operation and five min afterward. Figures illustrate the concentration for the particle size range of 0.5 to 4 µm, while the larger size concentration was negligible. The color scale defines number concentrations from 0 (blue) to 200 #/cm$^3$ (red). The values between these limits are mapped by blue, green, yellow, and orange. The purple shows values greater than 200 #/cm$^3$. As expected, the number concentration distribution varies with the operating conditions. For all scenarios, the smaller the particle size, the higher concentration is.

In closed-door scenarios, by comparing the scenario that no air purifier is running (Figure 3.a) with the scenario that the air purifier is running at the beginning of operation (Figure 3.b), it can be observed that particles have a wider distribution in Figure 3.b, which means particles are growing to the larger sizes. For instance, the concentration of higher than 200 #/cm$^3$ is observed for 0.5-1.3 $\mu$m particles in Figure 3.a, while, this range of concentration is observed for 0.5-1.5 $\mu$m particles in Figure 3.b. Moreover, the concentration of 200-70 #/cm$^3$ is detected for 1.3-2.8 $\mu$m particles in Figure 3.a; however, 1.5–3.5 $\mu$m particles have this concentration range in Figure 3.b. The real generated values for Figure 3.b are even more than this reported number because the removal process is started ad the beginning.

From this observation, it can be inferred that running the air purifier from the beginning causes air circulation in the room. The air circulation can enhance the interaction between airborne particles leading to agglomeration in the area that particles are generated [36]. Thus, the particles may grow to the larger ones when the air purifier was on at the beginning of the operation. Growing to larger sizes is preferable in terms of particle removal. Removal by HEPA filter is size dependant; the larger sizes, the more probable filtration is. The filtration of micron particles is due to interception and impaction [37].

Similar behavior was observed when the door was open. Comparing Figures 3.c with 3.d shows that growing particles to larger sizes during the first 5 min while the air purifier was running from the beginning of the operation. The concentration of higher than 200 #/cm$^3$ is observed for 0.5-1 $\mu$m particles in Figure 3.a (air purifier off), while, this range of concentration is observed for 0.5-1.4 $\mu$m particles in Figure 3.d. Moreover, the concentration of 200-70 #/cm$^3$ is detected for 1-2.2 $\mu$m particles in Figure 3.c, however 1.4–2.5 $\mu$m particles have this concentration range in Figure 3.d.

The particles generated in the 5-min long operation gradually spread in the room, and their concentrations were decreased by different mechanisms. They are introduced in the next sections.

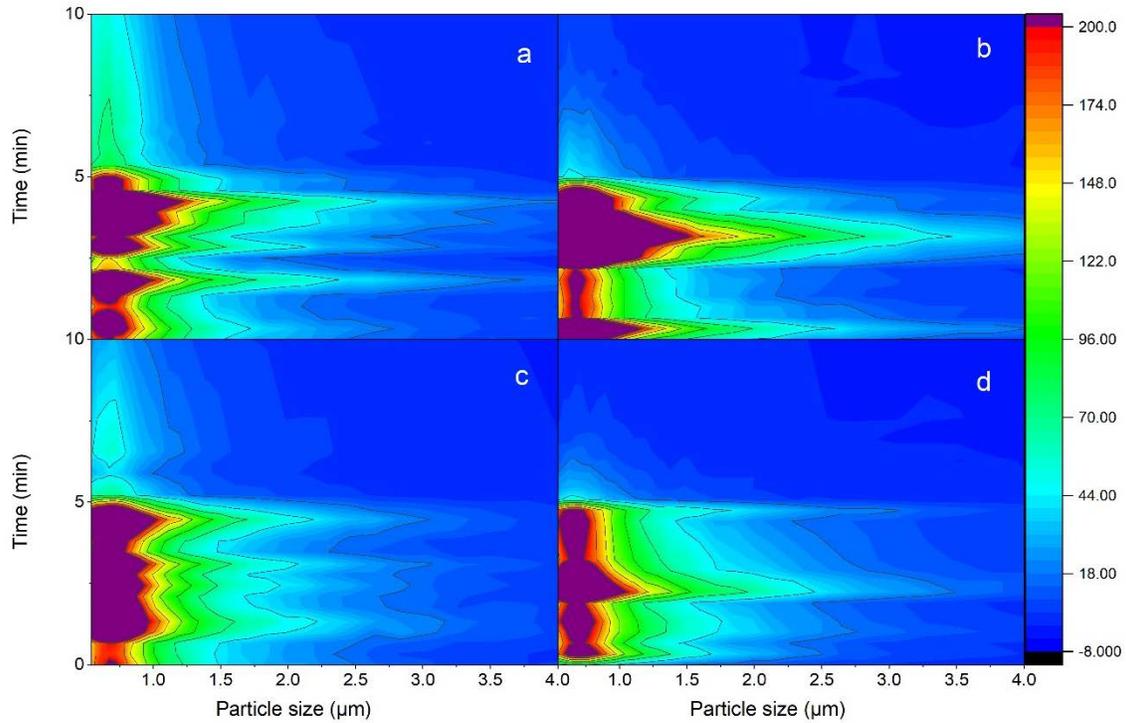

Figure 3. Concentrations of particles from 0.5 to 4 *μ*m in the first 10 min measurement with (a, and b) closed-door: (a) air purifier off, (b) high-speed air purifier turned on from the beginning of particle generation and with (c, and d) open-door, and c) air purifier off, d) high-speed air purifier burned on from the beginning of particle generation.

### 3.2. Spatial and temporal change of particle concentrations in the generated zone

*3.2.1. Effects of air purifier and the door condition on the spread and removal of 0.5-μm particles*

Figure 4 shows real-time number concentrations of 0.5 *μ*m particles during the dental operation and afterward until they reached the background level. Figure 4a is for the closed-door and Figure 4b is for the open-door scenarios. The solid horizontal line marks the background concentration of 0.5 *μ*m particles. The particle concentrations dropped gradually, likely by settlement on the surface [38], filtration by the air purifier, or dispersion in and out of the room. Table 3 summarizes the times it takes for the number concentrations to reach their background levels (removal times) for all six scenarios. In the worst-scenario scenario, when the door is closed and no air purifier is running in the room, it takes 95 min for 0.5 $\mu$m particles to return to the background level.

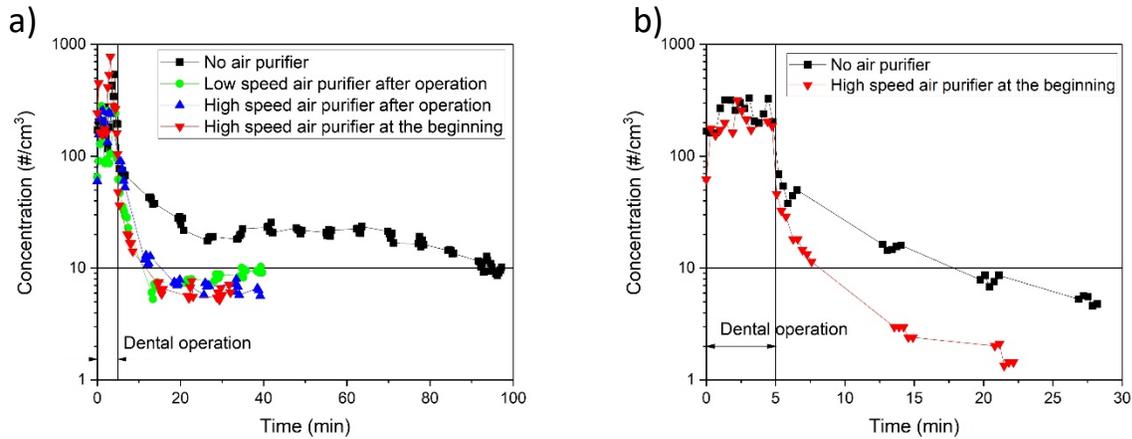

Figure 4. Spread and removal of 0.5-$\mu$m particle in (a) close door and (b) open door scenarios

Two conclusions can be drawn from the results in Figure 4. First, both Figures 4a and 4b show that the air purifier expedited particle removal from the air. For instance, Figure 4a shows that running high-speed air purifier enhanced the removal time of 0.5-$\mu$m particles at least 6.3 times faster than the scenario with no air purifier. Figure 4a shows the lowest particle concentrations in the room when the high-speed air purifier is running from the beginning of the operation. However, the removal time is almost the same for all these 3 scenarios: low-speed air purification after the dental operation, high-speed air purification after the dental operation, and high-speed air purification from the beginning of the operation. It can be inferred that particles were captured with the HEPA filter and Activated Carbon Filter installed in the air purifier. In addition to filtration, enhancing air circulation in the room by the air purifier leads to faster particle settlement on the surface areas. These results suggest that air purifier has a crucial role in removing airborne contamination of dental offices in the generation zone.

Second, comparing the removal times of open-door scenarios (Figure 4.b) with closed-door scenarios (Figure 4.a) shows that the open door expedited the removal of 0.5 $\mu$m particles in the generation zone. The open door enables the dispersion of airborne particles by natural ventilation and air circulation. Dispersed particles may settle on the indoor surfaces and exit the room. It implies that the number concentration in the hallway was lower than inside the test room at the time of these measurements. On the other hand, external particles may enter the room and worsen the inside air quality if there are more particles outside of the door. This was the scenario on another day of measurement (see supplementary information) when the dentist was operating on patients. Therefore, the opening window, similar to the open-door scenarios, is recommended as a short term solution for the dental offices without air filtration systems.

The particle removal time varies with particle size although the air purifier and open door help reduce the concentration of all-size particles in the generation zone. The next section elaborates on the size dependency of particle spread and removal because smaller particles probably carry more infectious microorganisms because the concentration of smaller particles is higher than the larger ones.

*3.2.2. Effects of particles size on particle removal*

Figure 5 demonstrates the number concentrations of particles with sizes of sub-0.5, 0.5, 1, and 2.5 $\mu$m for all six scenarios. The background concentration is shown with the red horizontal line for <0.5 and 0.5 $\mu$m particles, green dash line for 1 $\mu$m particles and blue dotted line for 2.5 $\mu$m particles. The removal time for different particle sizes is marked with asterisks and their corresponding values are listed in Table 3.

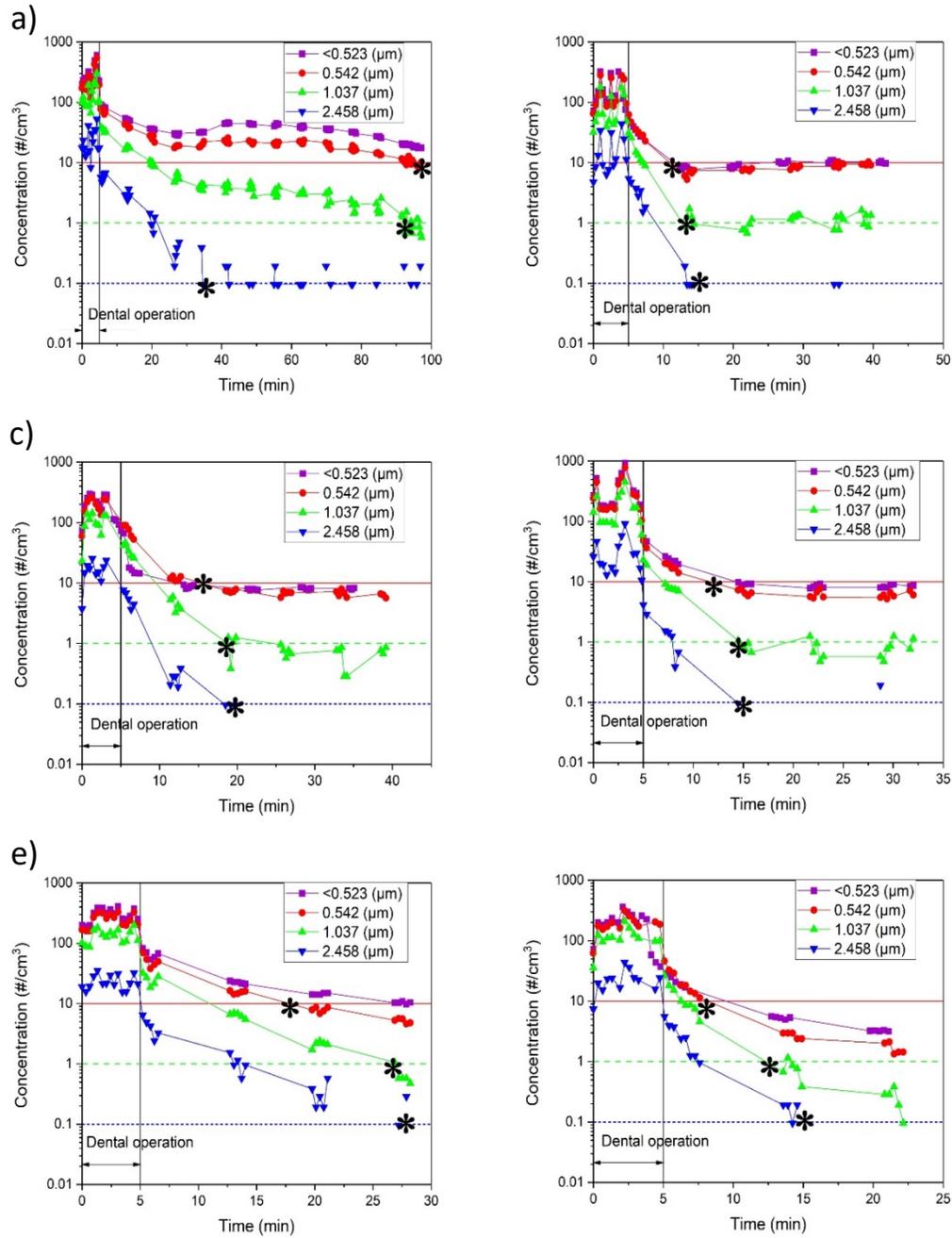

Figure 5. Number concentrations of sub-0.5, 0.5, 1, 2.5 $\mu$m particles measured in the generation zone for (a-d) closed-door scenarios: (a) air purifier off, (b) low-speed air purifier turned on after particle generation, (c) high-speed air purifier turned on after particle generation, (d) high-speed air purifier running from the beginning of particle generation and (e-f) open-door scenarios: (e) air purifier off, (f) high-speed air purifier running from the beginning of particle generation.

Table 2. Removal times of the scenarios at the generation zone

| No. | Door | Air purifier | | | Removal time at generation zone (min) | | |
|---|---|---|---|---|---|---|---|
| | | On/Off | Fan mode (Flow rate) | Air cleaning starting time | 0.5 $\mu$m | 1 $\mu$m | 2.5 $\mu$m |
| 1 | Close | Off | - | - | 95 | 92 | 35 |
| 2 | Close | On | Low (153 CFM) | After dental operation | 11 | 13 | 15 |
| 3 | Close | On | High (312 CFM) | After dental operation | 15 | 18.5 | 20 |
| 4 | Close | On | High (312 CFM) | At the beginning of the dental operation | 12 | 14.5 | 15 |
| 5 | Open | Off | - | - | 18 | 26.5 | 28 |
| 6 | Open | On | High (312 CFM) | At the beginning of the dental operation | 8 | 12.5 | 15 |

There are several mechanisms of particle removal from the air including settling, air circulation, and air filtration. First, all particles in a closed-door room without air circulation or filtration settle down because of gravity. It is well-known that the larger particles have higher gravitational settling velocity and that their removal times are shorter than the smaller particles. Figure 5a further confirms this mechanism. For example, 2.5-$\mu$m particles disappeared faster than those that were smaller. Second, air circulation leads to the dispersion of particles and their subsequent removal by settling on the surface areas or exiting the room or both. The drag force on a particle is also size-dependent. It usually takes a longer time for a larger particle to disperse than the smaller ones do. Figure 5e indicates that air circulation through the open door expedited the particle removal, although the air purifier was off. In addition, Figure 5e shows expedited removal of smaller particles and confirms that air circulation is the dominant mechanism in this scenario. Third, the filtration efficiency is also size dependant and it increased with the particle size for micron particles [37].

Figure 5.f for the high-speed air purifier running from the beginning of operation in the open door room shows the combined effects of all the three mechanisms. Air circulation may be the dominant mechanism of removal, although filtration also plays a significant role in the removal because it took a longer time to reduce the concentrations of 2.5-$\mu$m particles than the smaller ones.

Moreover, Figure 5.b, 5.c, and 5.d show that the removal times do not vary with particle size. Therefore, a combination of settling, air circulation, and air filtration all play roles in particle removal for these scenarios. Comparing these scenarios with that in Figure 5f demonstrates the strong effects of air circulation due to the open door.

In summary, an air purifier running at high fan speed may ensure the removal of 0.5 to 3 $\mu$m particles, while air circulation is more effective for smaller particles. Since the door of dental offices might be open frequently, an air purifier with a strong fan may help prevent cross-contamination from one room to the other through the door. Nonetheless, our study herein does not undermine the effectiveness of external high-volume evacuation (EHVE) and suction, which are often used near to the generation zone.

However, it does not mean that the room is completely cleaned even when the particle concentrations in the generation zone dropped back to the background. The particles may be transported from the source to the rest of the room. Dental staff walks around in the same room, and they often remove their masks for a short break at the corner, where there is little air circulation. It is necessary to investigate the spread of particles by analyzing the concentration at the corner of the room, and the results are presented in the next section.

## 3.3. Spatial and temporal change of particle concentrations in the corner of the dental office

*3.3.1. Effects of air purifier and the door condition on the spread and removal of 0.5-μm particles*

Figure 6 compares the number concentrations of 0.5-*μ*m particles in the corner with those at the generation zone for all 6 scenarios. This comparison helps quantify the number of particles in the corner when the number concentration in the generation zone dropped to the background level. The particles moved from the generation zone to the corner for some scenarios. Table 4 summarizes the times of travel indicated by the peaks and the ratio of concentrations in the corner to those in the generation zone. For example, the concentration peaks are observed for all sizes in 6 minutes when the door was closed and the air purifier was running. In this scenario, the number concentration of peak in the corner is lower than the value in the generation zone. On the contrary, Figure 6d and 6f show that no peak is observed in the corner for 0.5 *μ*m particles when the air purifier is running from the beginning of operation with either open or closed door. These results indicate the effectiveness of high-speed high-efficiency air purification. Generally, it can be inferred that the peak is observed in the corner when the rate of particle settlement and removal from the air is lower than particle transport to the corner.

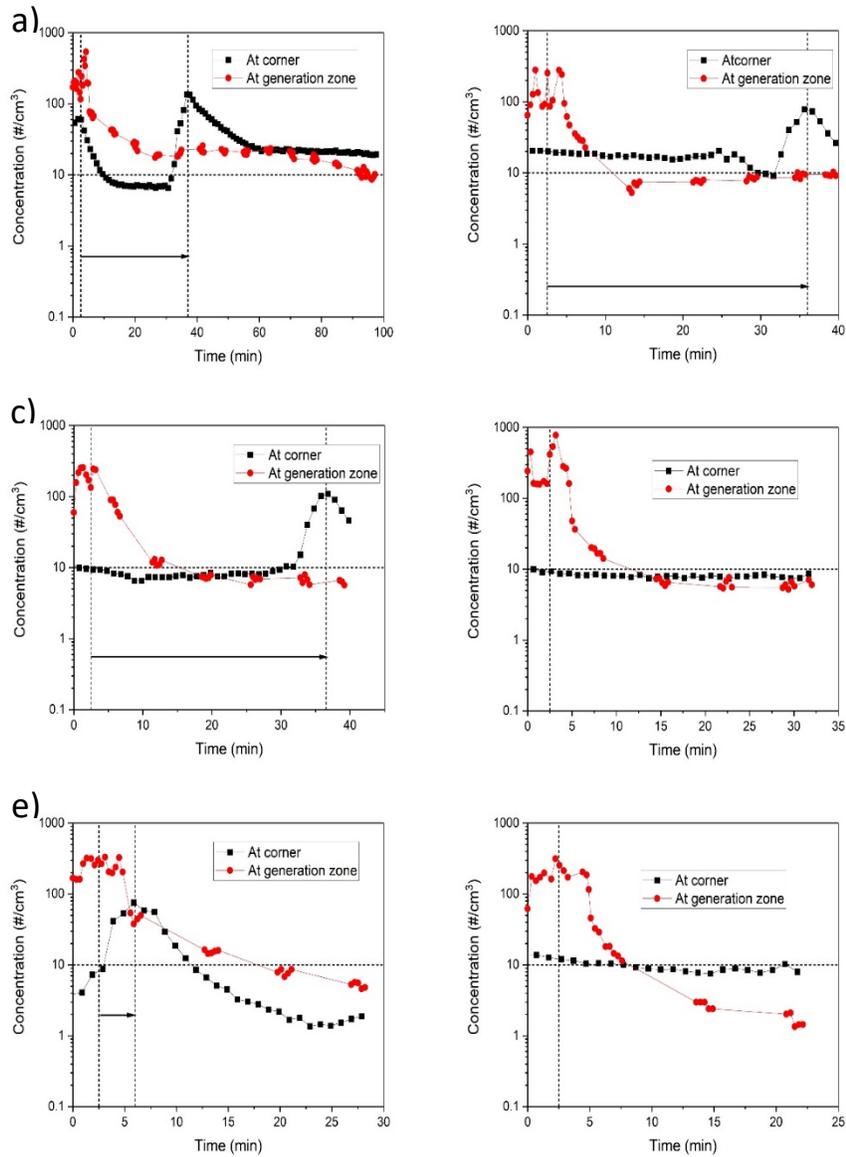

**Figure 6.** Comparison of the number concentrations of 0.5-$\mu$m particles in the corner with those at the generation zone for (a-d) closed-door scenarios: (a) air purifier off, (b) low-speed air purifier turned on after particle generation, (c) high-speed air purifier turned on after particle generation, (d) high-speed air purifier running from the beginning of particle generation and (e-f) open-door scenarios: (e) air purifier off, (f) high-speed air purifier running from the beginning of particle generation.

Table 3. The travel time and concentration ratios of 6 scenarios at the corner of the dental office

| No. | Door | Air purifier | | | Travel time (min) | | | Concentration ratio | | |
|---|---|---|---|---|---|---|---|---|---|---|
| | | On/Off | Fan mode (Flow rate) | Air cleaning starting time | 0.5 $\mu$m | 1 $\mu$m | 2.5 $\mu$m | 0.5 $\mu$m | 1 $\mu$m | 2.5 $\mu$m |
| 1 | Close | Off | - | - | 37 | 37 | 37 | 0.5 | 0.16 | 0.66 |
| 2 | Close | On | Low (153 CFM) | After dental operation | 36 | 36 | 36 | 0.33 | 0.1 | 0.4 |
| 3 | Close | On | High (312 CFM) | After dental operation | 36.5 | 36.5 | 36.5 | 0.33 | 0.11 | 0.5 |
| 4 | Close | On | High (312 CFM) | At the beginning of the dental operation | - | - | - | - | - | - |
| 5 | Open | Off | - | - | 6 | 6 | 6 | 0.26 | 0.11 | 0.5 |
| 6 | Open | On | High (312 CFM) | At the beginning of the dental operation | - | 21 | 21 | - | 0.016 | 0.06 |

Table 4 indicates that it took 6 min for the concentration peak to reach the corner when the door was open and air purifier off. In comparison, Figure 6a shows that the travel time is shorter when the door was closed with the air purifier off (37 min). The air circulation resulting from the open door affected the contamination level in the corner. Therefore, an open door during operation may expedite the travel of particles from the source to the corner.

The travel time of the concentration peak and peak concentration ratios are close to each other for the three closed-door scenarios including air purifier off (Figure 6a), low-speed air purifier running after the operation (Figure 6b), and high-speed air purifier running after the operation (Figure 6c). Thus, the same fraction of particles reaches the corner at the same time for these scenarios. This is surprising because these results imply that the air circulation result from the air purifier has little impact on the air movement to the corner of the room (*See* Figure 1).

*3.3.2. Effects of particles size on particle removal*

Figure 7 shows the number concentrations of 0.3-, 0.5-, 1-, 2.5-*μ*m particles in the corner of the office for all six scenarios. All particles reached the corner with the same travel time as indicated by the concentration peaks observed in the corner except one scenario. Figure 7f shows the concentration peaks for 1- and 2.5-*μ*m particles, but not for the 0.3- and 0.5-*μ*m particles. This observation is expected based on two conclusions that were made in section 3.2.2 and 3.3.1 for this scenario. First, the removal rate of larger particles is lower than the smaller ones while the air circulation due to the open door and filtration are removal mechanisms. Second, the peak is observed in the corner when the rate of particle settlement and removal from the air is lower than particle transport to the corner. Thus, a fraction of 1, 2.5 *μ*m particles, which is not removed from the air, traveled to the corner.

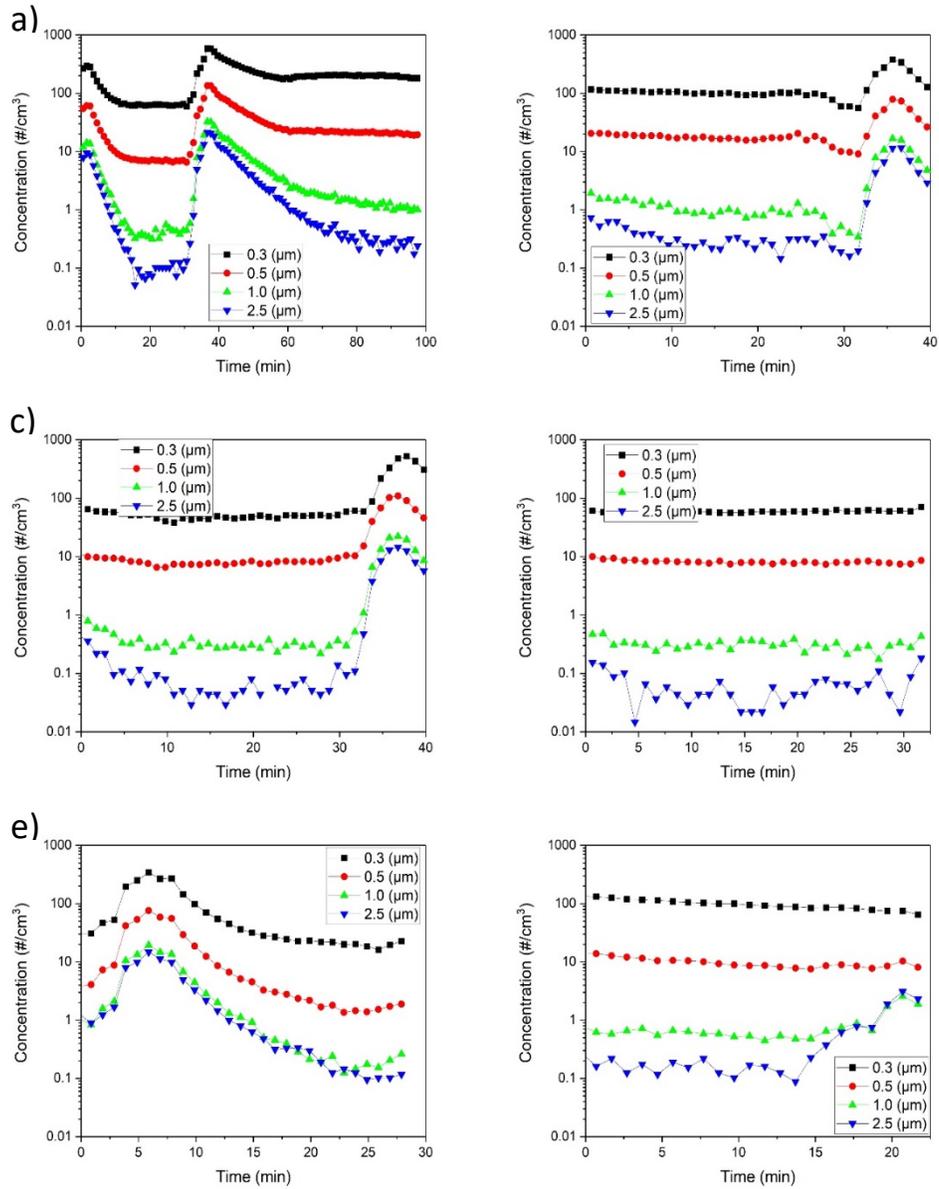

Figure 7. Number concentrations of sub-0.5, 0.5, 1, and 2.5 *µ*m particles measured in the corner of the office for (a-d) closed-door scenarios: (a) air purifier off, (b) low-speed air purifier turned on after particle generation, (c) high-speed air purifier turned on after particle generation, (d) high-speed air purifier running from the beginning of particle generation and (e-f) open-door scenarios: (e) air purifier off, (f) high-speed air purifier running from the beginning of particle generation.

## 4. Conclusions

The following conclusions can be drawn from the results of this study:

- In the worst-scenario scenario with no protection system in the closed-door office and continuous high-speed drilling, it takes 95 min for 0.5 $\mu$m particles to return to background level and that it takes a shorter time for particles larger than 0.5 $\mu$m to be removed from the air. In the real operations with the patient, which usually is less than five minutes, air may be cleaner because of other measures like suction from the source (*i.e.*, the mouth).

- There are three size-dependent mechanisms for particle removal: gravity settling, air circulation, and air filtration. Technologies that combine all of them are the most effective in air cleaning. The air purifier expedited the removal time at least 6.3 times faster than the scenario with no air purifier in the generation zone. Running high-speed air purifier at the beginning of the operation is the most effective scenario in reducing airborne particle concentrations. The air purifier at one corner could not eliminate the concentration peak in the other corner of the room except for the scenario when the door was closed and the air purifier was running at the highest speed from the beginning of the operation.

- It is recommended to keep the door closed during the operation; otherwise, particles may enter the hallway through the open door. These particles may transmit diseases if they carry infectious microorganisms. In the dental offices without air purification devices, it is recommended to open the window(s) when possible to promote natural ventilation; however, it may cause accumulation of particles in a corner. In addition, staff should leave the room after the operation and close the door for particles to settle or exit the window(s). Admittedly, the surfaces should be cleaned where particles may settle on.

- Our results have important implications for infectious disease transmission in closed settings such as dentists and doctors' offices. Although we did not study other closed environments such as schools, our study documents the time taken for airborne particles to settle down as well as the utility of air purifiers, which highlights the importance of air circulation and filtration in closed settings. In the context of the current COVID-19 pandemic, our study findings can assist in developing guidelines for air circulation and filtration, which can significantly reduce the chances of disease transmission.

## Acknowledgments

The authors would like to acknowledge the financial and technical support from the Natural Sciences and Engineering Research Council of Canada (RGPIN-2020-04687), the GCI ventures Capital, Surgically Clean Air, and Waterloo Filtration Institute.

# Supplementary Material for

**In situ Measurement of Airborne Particle Concentration in a Real Dental Office: Implications for Disease Transmission**

Maryam Ravazi[1], Zahid Butt[2], Mark H.E. Lin[3], Helen Chen[2], Zhongchao Tan[1*]

[1] Department of Mechanical and Mechatronics Engineering, University of Waterloo, Ontario, Canada
[2] School of Public Health and Health Systems, University of Waterloo
[3] The Institute for Dental Excellence Inc., 88 Finch Ave, East North York, Ontario, Canada

**Study design of real operation with the patient**

In the second part of this study, the number concentrations were measured for three dental operations with real patients. The air ventilation system was blocked, and the door was closed, however, it was opened several times during the operations.

Table S- 1 summarizes the conditions for these three operations. The first and second operations were done in the large room 4 × 5 × 3 m (W × L × H). The sampling tube was 30 cm away from the patient's head in the left hands side of the doctor's chair. The sampling tube was blocked several times by the doctor's arm during the operation. In the first operation, the air purifier was running in periodic mode between low speed (153 CFM) and turbo speed (406 CFM), while it was off for second and third operations. The third operation was conducted in a small room. The sampling tube was 30 cm away from the patient's head. The sampling tube was located on the opposite side of the doctor to avoid blockage.

**Table S- 1.** Dental operation conditions

|  | Dental operation | | Room size | Air purifier | Room condition before the operation | | Room condition after operation | |
| --- | --- | --- | --- | --- | --- | --- | --- | --- |
|  | Type of operation | duration | | | Temperature ($^0$C) | Humidity (%) | Temperature ($^0$C) | Humidity (%) |
| Operation 1 | Filling with 1 root canal High/low-speed handpieces | 88 min | Large (Room A) | On | 21.9 | 60 | 23.6 | 58 |
| Operation 2 | Filling with 1 root canal High/low-speed handpieces | 33 min | Large (Room A) | Off | 24.7 | 54 | 24.9 | 49 |
| Operation 3 | Crown insertion High/low-speed handpieces | 73 min | Small (Room B) | Off | 23.3 | 51 | 24.3 | 51 |

**Field campaign with real patients**

Figure S- 1, Figure S- 2, and Figure S- 3 demonstrate the number concentration of <0.5, 0.5, 1, 2.5, and 5 $\mu$m particles during three dental procedures. The horizontal lines mark the background concentrations of <0.5 and 0.5 $\mu$m particles. The moments that door was open are shown with Asterisks in the fig, which last less than a minute.

The first operation was conducted in 2 parts, shown in by patterned area. The higher concentrations entered the room from outside. After closing the door, the number concentration was reduced by the air purifier. Moreover, the concentration peaks were observed, in the moments that the door was open. The major fraction of particles was generated in the second part of the operation. During this time, the air purifier was running at low speed in 7 min and turbo speed in 7 min. In the first 7 min, the removal rate was 0.28 (#/cm$^3$min) and the second 7 min was 1.14 (#/cm$^3$min), 4 times faster than the time with low speed.

The second operation was conducted in a single part, and no considerable particles were measured. Similar to the fist operation, the number concentration of outside was higher than inside. The number concentration in the third operation was higher than the first two operations. The third operation was conducted in 2 parts. Higher values of concentration coming from outside are observed in this operation comparing to the first two because APS was closer to the door in 3$^{rd}$ operation.

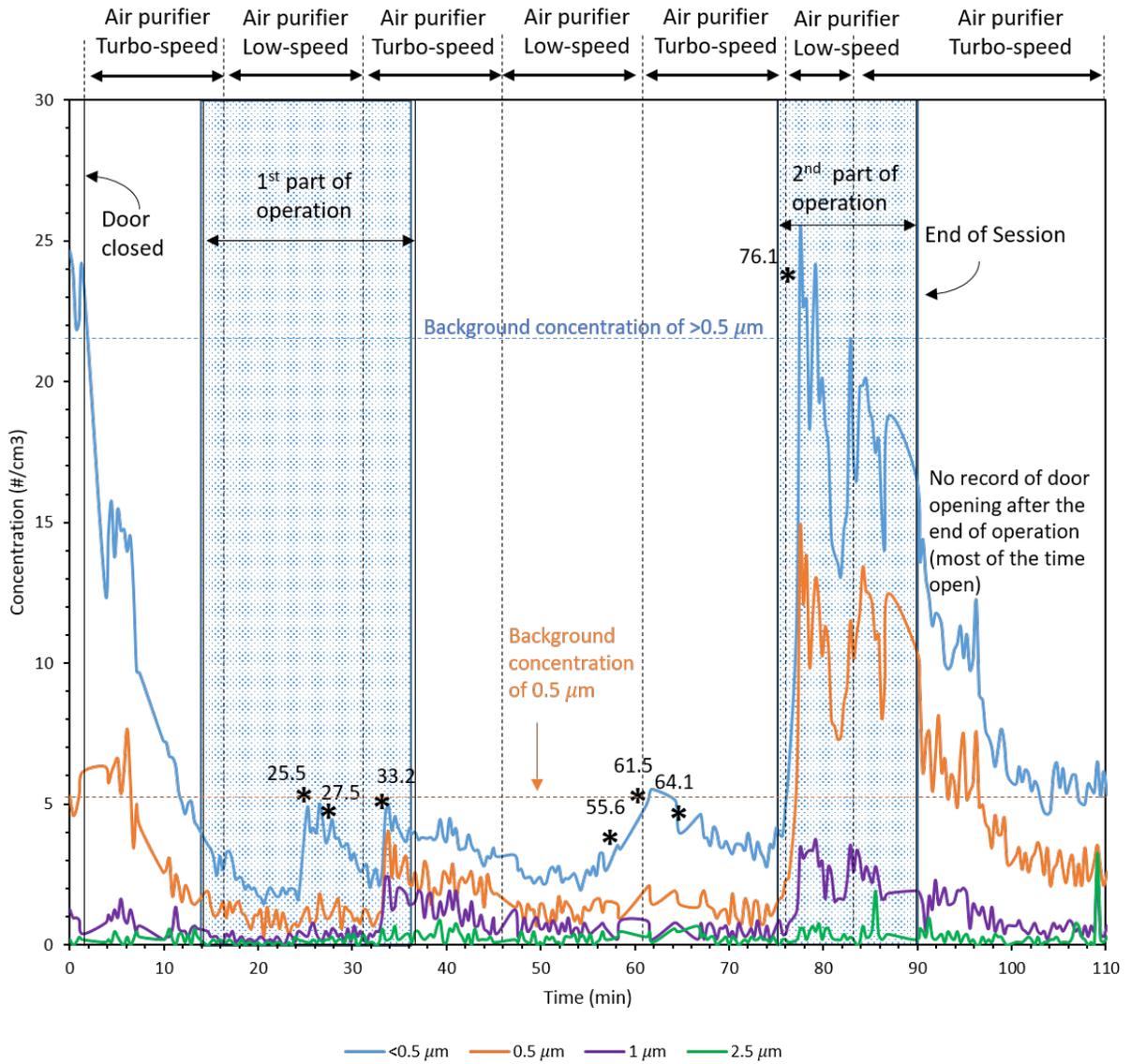

**Figure S-1** Operation 1 in room A (Large room)

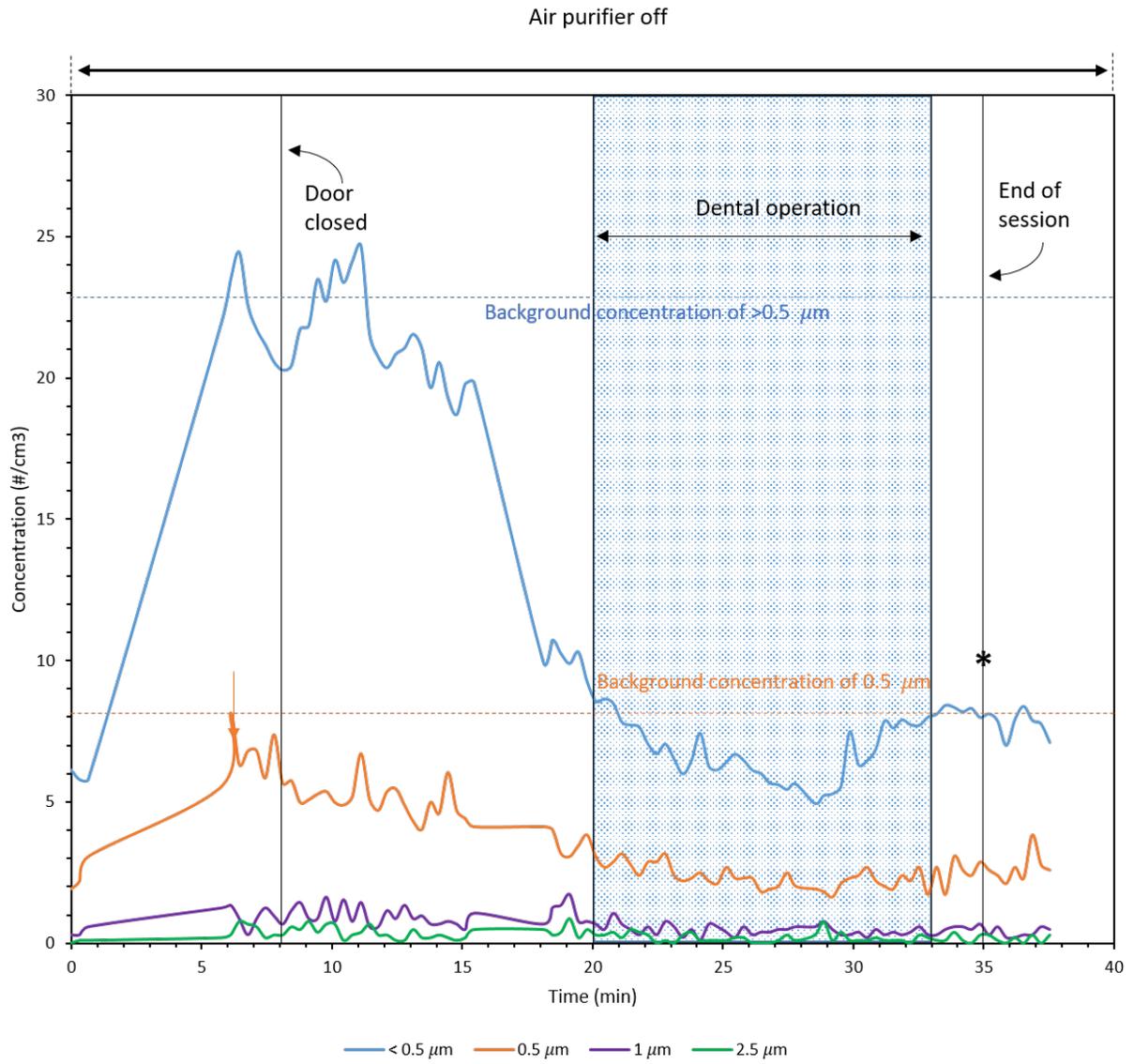

**Figure S- 2** Operation 2 in room A (Large room)

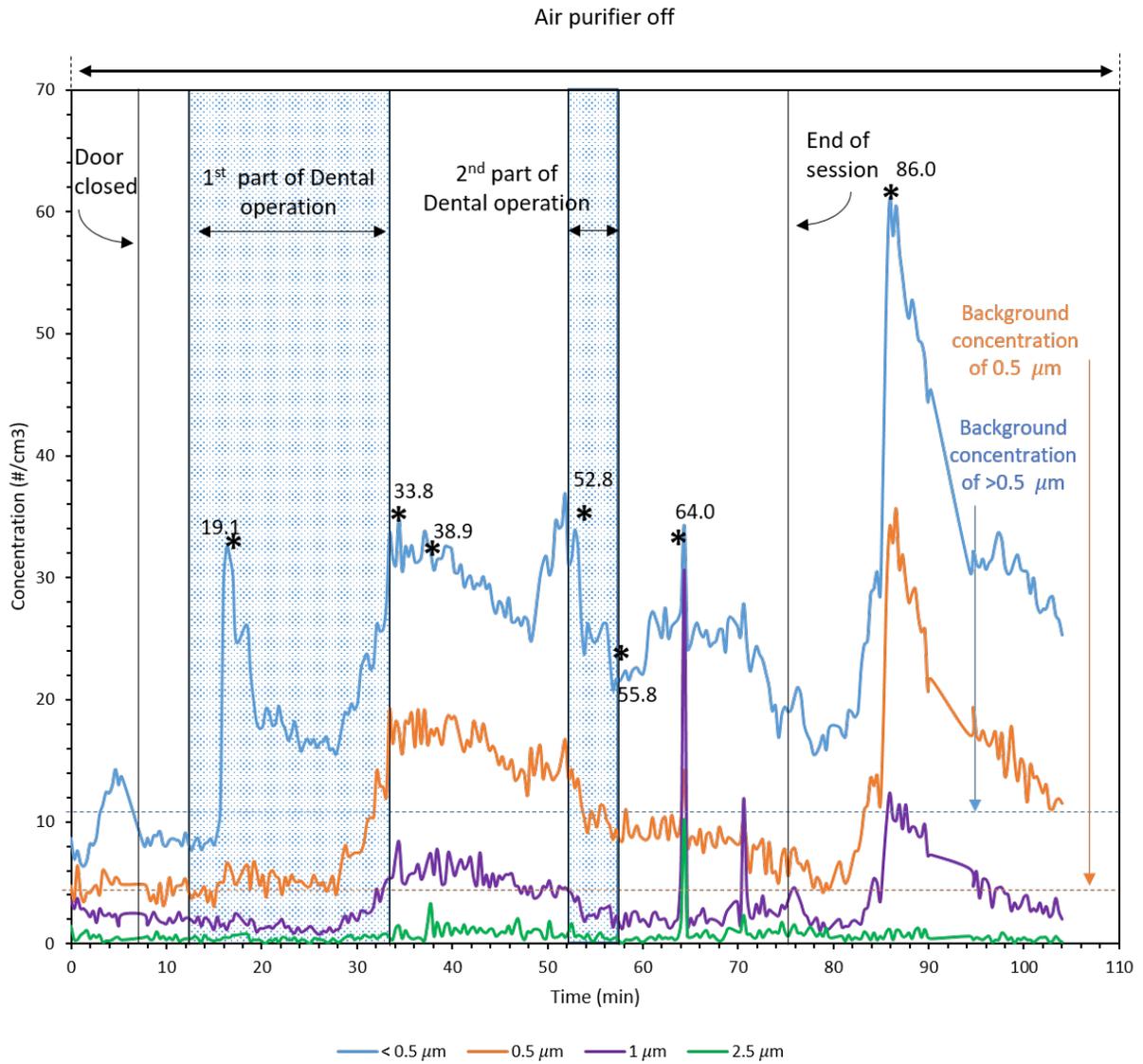

**Figure S- 3 Operation 3 in room B (small room)**